\documentclass[a4paper,12pt]{article}
\usepackage[T2A]{fontenc}
\usepackage[cp1251]{inputenc}
\usepackage{graphicx}
\usepackage{epsf}
\usepackage{epsfig}
\usepackage{amssymb,amsthm,amsmath}
\usepackage{latexsym}
\input{epsf}
\newcommand{\be}{\begin{eqnarray}}
\newcommand{\ee}{\end{eqnarray}}
\newcommand{\ba}{\begin{array}}
\newcommand{\ea}{\end{array}}

\begin{document}
\renewcommand{\thefootnote}{\fnsymbol{footnote}}
%
%

%
\begin{center}
{\Large Dual parameterization and Abel transform tomography\\  for twist-3 DVCS}\\
\vspace{0.35cm}
 Alena M. Moiseeva$^{a,b}$ and Maxim V. Polyakov$^{b,c}$\\

\vspace{0.35cm}
 $^a$ Bogoliubov Laboratory of Theoretical Physics,
  JINR, 141980, Moscow Region, Dubna, Russia\\
$^b$Institut f\"ur Theoretische Physik II,
Ruhr--Universit\"at Bochum, D--44780 Bochum, Germany\\
$^c$Petersburg Nuclear Physics
Institute, Gatchina, St.\ Petersburg 188350, Russia

%
%
\end{center}

\begin{abstract}
We derive dual parameterization for the DVCS amplitude to the
twist-3 accuracy. The relation of the dual parameterization to the
Abel transform tomography for amplitudes of hard exclusive
processes is elaborated. We prove dispersion relations for the
twist-3 DVCS amplitude in the Wandzura -- Wilczek approximation
and show that the subtraction constant is given by the D-term form
factor. The same as in the case of twist-2 DVCS amplitude.
\end{abstract}
\vspace{0.1cm}

\section*{\normalsize \bf Dual parameterization facilitates  Abel tomography for DVCS amplitudes}
\noindent The dual parameterization \cite{MaxAndrei} of
generalized parton distributions (GPDs) \cite{pioneers} (for
recent reviews of GPDs see \cite{GPV,Diehlrev,Belitskyrev,Boffi})
provides us handy and flexible tool to describe amplitudes of hard
exclusive processes in terms of single function -- so-called
quintessence function. This function quantifies the maximum amount
of information about GPDs, which can be extracted from data on
hard exclusive processes. First phenomenological applications of
this parameterization to DVCS give very promising results
\cite{Guzey,tamed}. In present paper we extend the dual
parameterization to the twist-3 GPDs in the Wandzura--Wilczek
approximation.

Let us first remind basics of the dual parameterization and dwell on its relation to Abel tomography
\cite{Abel}.
The leading order twist-2 amplitude of hard exclusive reactions is
expressed in terms of the following elementary
amplitude\footnote{As the first step we restrict ourselves to DVCS on spinless hadron.}:

 \be A^{{\rm tw2}}(\xi,t)=\int_{0}^1 dx\,H(x,\xi,t)\, \left[\frac
1{\xi-x-i0} -\frac 1{\xi+x-i0}\right]. \label{elementaryAMP} \ee
We see that the amplitude is given by the convolution integral in
which dependence of GPDs on variable $x$ is ``integrated out''.
One can
not completely restore the GPD $H(x,\xi,t)$ from Eq.~(\ref{elementaryAMP}) , because from this equation
one can obtain the same amplitude with different GPDs $H(x,\xi,t)$. In dual parameterization
the expression for the amplitude has the form \cite{MaxAndrei,tomography,educing}:
\begin{eqnarray}
{\rm Im\ } A^{{\rm tw2}}(\xi,t)&=&
\int_{\frac{1-\sqrt{1-\xi^2}}{\xi}}^1 \frac{dx}{x} N(x,t)\ \Biggl[
\frac{1}{\sqrt{\frac{2 x}{\xi}-x^2-1}}
\Biggr]\, ,
\label{IM}
\\
\nonumber {\rm Re\ } A^{{\rm tw2}}(\xi,t)&=&
\int_0^{\frac{1-\sqrt{1-\xi^2}}{\xi}} \frac{dx}{x} N(x,t)\ \Biggl[
\frac{1}{\sqrt{1-\frac{2 x}{\xi}+x^2}} + \frac{1}{\sqrt{1+\frac{2
x}{\xi}+x^2}}-\frac{2}{\sqrt{1+x^2}}
\Biggr]  \\
&+&\int^1_{\frac{1-\sqrt{1-\xi^2}}{\xi}} \frac{dx}{x} N(x,t)\
\Biggl[ \frac{1}{\sqrt{1+\frac{2
x}{\xi}+x^2}}-\frac{2}{\sqrt{1+x^2}} \Biggr]+ 2 D(t) \, .
\label{RE}
\end{eqnarray}
The amplitude is completely determined by, so-called, GPD-quintessence function $N(x,t)$ \cite{tamed}
and by the D-form factor $D(t)$. The latter is : \be D(t)=\sum_{n=1}^\infty
d_n(t)=\frac 12 \int_{-1}^1 dz\ \frac{D(z,t)}{1-z}\, ,
\label{DFF} \ee where
$D(z,t)$ is the D-term \cite{PW99}. One can check
\cite{tomography} that the amplitude given by Eqs.~(\ref{IM},\ref{RE})
automatically satisfies the dispersion relation with the
subtraction constant given by the D- form factor, as it should be
on general grounds \cite{Teryaev,DiehlIvanov}. The GPD-quintessence function $N(x,t)$
is expressed in terms of set of forward-like function\footnote{
We call the functions
$Q_{2\nu}(x,t)$ {\it forward-like} because \cite{tomography}:

\begin{itemize}
\item
At the LO scale dependence of functions $Q_{2\nu}(x,t)$ is given
by the standard DGLAP evolution equation, so that these functions
behave as usual parton distributions under QCD evolution.

\item
The function $Q_0(x,t)$ is related to the forward $t$-dependent
quark densities $q(x,t)$ (at $t=0$ they are reduced to usual
parton densities $q(x)$ measured in DIS) as:

\be Q_0(x,t)= q(x,t)+\bar q(x,t)-\frac x2  \int_x^1
\frac{dz}{z^2}\ ( q(z,t)+\bar q(z,t)) \, . \label{Q0} \ee

\item
The expansion of the GPD $H(x,\xi,t)$ around the point $\xi=0$
with fixed $x$  to the order $\xi^{2\nu}$ involves only finite
number of functions $Q_{2\mu}(x,t)$ with $\mu\leq \nu$.
\end{itemize}
See original papers \cite{MaxAndrei,tomography,educing} for
detailed discussion of properties of these functions and their physics interpretation.} as:

\be N(x,t)=\sum_{\nu=0}^\infty x^{2\nu}\ Q_{2\nu}(x,t)\, .
\label{QF} \ee

\begin{figure}
\begin{center}
\includegraphics
[ scale=1.3
]%
{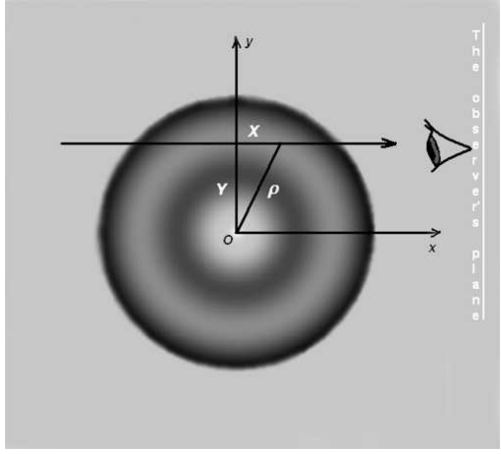}%
\end{center}
\caption{Illustration for Abel transform tomography} \label{foto}
\end{figure}

The knowledge of  the LO amplitude is in one-to-one equivalence to
the knowledge of the function $N(x,t)$ and the D-form factor
$D(t)$, because the Eq.~(\ref{IM}) can be inverted
\cite{tomography}, i.e. the function $N(x,t)$ can be expressed
{\it unambiguously} in terms of amplitude.
As the function $N(x,t)$ is unambiguously restored from the
amplitude it means that it contains maximal
information about GPDs which one can, in principle, obtain from the amplitudes.
This inversion,
actually, corresponds to the Abel transform tomography \cite{Abel}. Let us show relation  of Eq.~(\ref{IM}) to tomography\footnote{
\noindent
See Ref.~\cite{OlegRadon}
for the pioneering applications of the Radon tomography methods to GPDs.}. It is useful to make
the following substitution for the integration variable $x$ in Eq.~(\ref{IM}):
$$ \frac 1w = \frac 12\left(x+\frac 1x\right)\, .$$
This substitution correspond to famous Joukowskii conformal map \cite{Zhu},
which historically was used to understand some principles of aerofoil design.
After this change of variables the expression for the imaginary part of the amplitude gets the form:

\be
{\rm Im\ } A^{{\rm tw2}}(\xi,t)&=& \int_\xi^1 \frac{dw}{w}\ M(w,t)\ \frac{\sqrt \xi}{\sqrt{w -\xi}}\, ,
\label{imsimple}
\ee
where the function $M(w,t)$ is related to the function $N(x,t)$ as:

\be
M(w,t)=N\left( \frac{1-\sqrt{1-w^2}}{w},t\right)\ \frac{w}{\sqrt{2(1-w^2)}\sqrt{1-\sqrt{1-w^2}}}\, .
\ee
The Eq.~(\ref{imsimple}) is typical for the Abel tomography. In two dimensions, the Abel transform $a(y)$ can be interpreted as the projection of a
axially symmetric function $m(\rho)$ along a set of parallel lines of sight which are at distance $y$ from the origin. Referring to Fig.~\ref{foto}, the observer see
the image:
\be
a(y)=\int_{-\infty}^{\infty}dx\ m(\rho)\, .
\label{obraz}
\ee
Now, using axial symmetry of the function $m(\rho)$, we make
substitution for the integration variable $x^2=\rho^2-y^2$, as the
result we obtain the following expression for the image function
$a(y)$: \be a(y)= \int_{y^2}^\infty d\rho^2\
\frac{m(\rho)}{\sqrt{\rho^2 -y^2}}\, . \ee This equation has
exactly the form (with obvious identification of variables and
functions) of Eq.~(\ref{imsimple}). We see that we reduce our
problem about GPDs to the restoring of images of axially symmetric
objects\footnote{It would be interesting to understand the deep
meaning of this axial symmetry in context of GPD problem.
Amplitudes of hard exclusive processes are photographs of some
object which is axial symmetric in plane of variables
$X=\sqrt{\frac{2 x}{1+x^2}-\xi}$ and $Y=\sqrt\xi$.} from their
photographs on plane. This problem is solved unambiguously with
help of Abel transform \cite{Abel}. Let us demonstrate us this
directly for Eq.~(\ref{imsimple}). Introducing the following
notations $m(w,t)=\frac{M(w,t)}{w}$ and
$a(\xi,t)=\frac{1}{\sqrt{\xi}} {\rm Im} A (\xi,t)$ ($\rho^2=w$ and
$y^2=\xi$), we arrive at the integral equation of Abel:
$$a(\xi,t)=\int^{1}_{\xi}dw\ \frac{m(w,t)}{\sqrt{w-\xi}},$$
which has very simple and unambiguous solution.
Indeed, to solve the Abel equation
we multiply both sides of the above equation equality  by $1/\sqrt{\xi-z}$
and integrate it over $\xi$ from $z$ to 1. Then changing sequence of
integration over $\xi$ and $w$, we obtain
$$\int^{1}_{z}d\xi\ \frac{a(\xi,t)}{\sqrt{\xi-z}}=
\int^{1}_{z}d\xi\ \frac{1}{\sqrt{\xi-z}}\int^{1}_{\xi}dw\ \frac{m(w,t)}{\sqrt{w-\xi}}=\int^{1}_{z}dw\ m(w,t)\int^{w}_{z}d\xi\ \frac{1}{\sqrt{(w-\xi)(\xi-z)}}$$
$$\int^{1}_{z}d\xi\ \frac{a(\xi,t)}{\sqrt{\xi-z}}=\pi\int^{1}_{z}dw\ m(w,t).$$
Now differentiating that expression over $z$, one can write
$$m(w,t)=-\frac{1}{\pi}\frac{d}{dw}\int^{1}_{w}d\xi\
\frac{a(\xi,t)}{\sqrt{\xi-w}}.$$
This solves the Abel tomography problem. Making necessary trivial change of variables we obtain
the expression for the GPD- quintessence function $N(x,t)$ in terms of the amplitude \cite{tomography}:

\be
N(x,t)=\frac{2}{\pi}\ \frac{x(1-x^2)}{~~~(1+x^2)^{3/2}}
\int_\frac{2 x}{1+x^2}^1\frac{d\xi}{\xi^{3/2}}\
\frac{1}{\sqrt{\xi-\frac{2 x}{1+x^2}}} \left\{ \frac 12{\rm Im\ }
A^{{\rm tw2}}(\xi,t)-\xi \frac{d}{d\xi}{\rm Im\ } A^{{\rm tw2}}(\xi,t) \right\}
\label{nina} \ee
The beauty of the Abel transform lies in its exactness and conciseness. Many effective numerical methods  to implement
the Abel transform were developed, see e.g \cite{Li}. For instance, Chan and Lu \cite{Chan} suggested the form of Abel transform
which avoids derivatives (which can be noisy when applied to data) of the ``image function'' (${\rm Im} A$ in our case).
Our Eq.~(\ref{nina}) in representation of Chan and Lu has the form:

\be
\nonumber
N(x,t)&=&-\frac{1}{\pi}\ \frac{x(1-x^2)}{~~~(1+x^2)^{3/2}}
\int_\frac{2 x}{1+x^2}^1 d\xi\
\frac{1}{(\xi-\frac{2 x}{1+x^2})^\frac{3}{2}} \\
\label{ninochka}
 &\times &\left\{ \frac{1}{\sqrt \xi}\ {\rm Im\ }
A^{{\rm tw2}}(\xi,t)-\sqrt{\frac{1+x^2}{2 x }}{\rm Im\ } A^{{\rm tw2}}\left(\frac{2 x}{1+x^2},t\right) \right\}\\
&+&\frac{1}{\pi}\ \frac{\sqrt{2 x}(1+x)}{\sqrt{1+x^2}}\ {\rm Im\ } A^{{\rm tw2}}\left(\frac{2 x}{1+x^2},t\right)\, .
\nonumber \ee

\section*{\normalsize \bf Dual parameterization for twist-3 DVCS amplitude}
\noindent
The DVCS twist-3 amplitude  was computed in
Refs.~\cite{Anikin,Penttinen,BM,RW1}, it includes the contributions which are suppressed by power of $1/Q$
relative to the twist-2 amplitude. The inclusion of such terms is mandatory to ensure
the electromagnetic gauge invariance of the DVCS amplitude to the
order $\sqrt{-t}/Q$, also these contributions can give sizable
contributions to the DVCS observables \cite{KPV}. At the order $1/Q$ the DVCS amplitude depends on a
set of new GPDs.
In Refs.~\cite{BM,KPST,RW1,KP} it was shown in
that in the so-called Wandzura-Wilczek (WW) \cite{WW}
approximation these new functions can be expressed in terms of
twist-2 GPDs.
The WW approximation corresponds to neglecting  the nucleon
matrix elements of the `genuine twist-3' operators of the type
$\bar \psi G \psi$, {\it i.e.} neglecting the nonperturbative
quark-gluon correlations in the nucleon. In principle this
approximation is not justified by any small parameter of the
theory. However, recent measurements of the polarized structure
function $g_T(x)$ \cite{g2izmer} show that the WW approximation
works pretty well in this case. Also the estimates of the `genuine
twist-3' contributions
polarized structure functions $g_T(x)$
and $h_L(x)$  \cite{BPW,DP}, and to twist-3 GPDs \cite{Kiptily} in the theory of instanton vacuum
(which is a model of non-perturbative quark-gluon correlations)
showed that these contributions are parametrically suppressed
relative to the `kinematical' part of the twist-3 contributions by
the packing fraction of instantons in the vacuum.

To the twist-3 accuracy the DVCS amplitude has the form:

\be
\label{T}
T^{\mu\nu} &=&
\frac{1}{2P\cdot Q}\int dx
\Biggl(
\frac{1}{\xi-x-i 0} - \frac{1}{\xi+x-i 0}
\Biggr)
\Biggl[
H(x,\xi)
\Biggl(
-2\xi P^{\mu}P^{\nu}  -
P^{\mu}Q^{\nu} - P^{\nu}Q^{\mu}
\Biggr.
\nonumber\\
\Biggl.
&+&g^{\mu\nu}(P\cdot Q) - \frac{1}{2}P^{\mu}\Delta^{\nu}_{\perp} +
\frac{1}{2}P^{\nu}\Delta^{\mu}_{\perp}
\Biggr)
\nonumber\\
&-&\bigl[ H_3(x,\xi)+\frac \xi x H_A(x,\xi)\bigr]
\Delta^{\nu}_\perp
\Biggl(
3\xi P^{\mu} + Q^{\mu}
\Biggr)\Biggr]\, ,
\ee
where $P=(p+p')/2, Q=(q+q')/2,
\Delta=q-q'=p'-p = -2 \xi P +\Delta_\perp$, while $p,p',q,q'$ are the
initial and final momenta of pion and photon, respectively.
The three terms in the square brackets are individually gauge invariant up
to order $\Delta_{\perp}^2$, i.e. $q'_{\nu}T^{\mu\nu}=O(\Delta_{\perp}^2)=
q_{\mu}T^{\mu\nu}$.
As $P$ and $Q$ are longitudinal
and  $\Delta_{\perp}$ transverse, the first term
corresponds to transverse polarization of the virtual photon, whereas the second term
to the longitudinally polarized virtual
photon
The second term contains the new
twist-3 contributions to the DVCS amplitude, which are defined as:

\be
\label{Parm}
\int \frac{d\lambda}{2\pi} e^{i\lambda x (\bar P n)}
\langle p'| \bar \psi\biggl(-\frac{\lambda}{2}n
\biggr) \gamma^\mu
\psi\biggl(\frac{\lambda}{2}n\biggr)|p\rangle&=&
P^\mu H(x,\xi) + \Delta^\mu_\perp H_3(x,\xi)\, , \\
\int \frac{d\lambda}{2\pi} e^{i\lambda x (\bar P n)}
\langle p'| \bar \psi\biggl(-\frac{\lambda}{2}n
\biggr) \gamma^\mu \gamma_5
\psi\biggl(\frac{\lambda}{2}n\biggr)|p\rangle&=&
i\varepsilon_{\mu \alpha \beta \delta}\Delta^{\alpha}P^{\beta}
n^{\delta}H_A(x,\xi) \, .
\ee
Here the light-cone vector $n$ is normalized as $n\cdot P=1$

We introduce elementary twist-3 amplitude (in complete analogy with twist-2 that (\ref{elementaryAMP})), which corresponds to longitudinally polarized
virtual photon:
\be
\label{elementaryTW3}
A^{{\rm tw3}}(\xi,t)=
 -\frac{\xi}{2}\int^{1}_{0}dx\left(\frac1{\xi-x-i0} -\frac 1{\xi+x-i0}\right)4\left[H_{3}(x,\xi,t)+\frac{\xi}{x}H_{A}(x,\xi,t)\right].
\ee
This amplitude can be extracted from the DVCS data considering azimuthal angle dependence of the observables.

For the new twist-3 GPDs we use WW approximation \cite{KPST,Anikin:2001ge}:

 \begin{eqnarray}
 4\left[H_{3}(x,\xi,t)+\frac{\xi}{x}H_{A}(x,\xi,t)\right]&=&
 \frac{x-\xi}{x} \left\{\theta(\xi-x)\int^{x}_{-1}\frac{dy}{y-\xi}\partial_{+}H-
 \theta(x-\xi)\int^{1}_{x}\frac{dy}{y-\xi}\partial_{+}H\right\}
 \nonumber\\
 &+& \frac{x+\xi}{x}\left\{\theta(-x-\xi)\int^{x}_{-1}\frac{dy}{y+\xi}\partial_{-}H
 -\theta(x+\xi)\int^{1}_{x}\frac{dy}{y+\xi}\partial_{-}H \right\},
 \nonumber
 \end{eqnarray}
 where
 $$\partial_{\pm}H = \frac{\partial H(x,\xi,t)}{\partial\xi}\pm\frac{\partial H(x,\xi,t)}{\partial x}.$$
After substitution of this expression into
Eq.~(\ref{elementaryTW3}) we obtain the expression for the twist-3
amplitude in terms of twist-2 GPD $H(x,\xi,t)$: \be
\label{elementaryTW3WW} A^{{\rm tw3}}(\xi,t)=\frac{\xi}{2}\int^{1}_{0}dx\frac{2x}{x^{2}-\xi^{2}}\left[\theta(\xi-x)\frac{x-\xi}{x}\int^{x}_{-1}\frac{dy}{y-\xi}\partial_{+}H-
\theta(x+\xi)\frac{x+\xi}{x}\int^{1}_{x}\frac{dy}{y+\xi}\partial_{-}H\right].
\ee

To compute the twist-3 amplitude we  employ formal series decomposition of twist-2 GPD $H(x,\xi,t)$, which represents GPDs as
the infinite sum of t-channel resonance exchanges \cite{MVP98}:
\be
H(x,\xi,t)=\sum^{\infty}_{n=1,3...}\sum^{n+1}_{l=0,2...}B_{nl}(t)\theta\left(1-\frac{x^{2}}{\xi^{2}}\right)\left(1-\frac{x^{2}}{\xi^{2}}\right)
C^{3/2}_{n}\left(\frac{x}{\xi}\right)P_{l}\left(\frac{1}{\xi}\right).
\label{formal}
\ee
It is easy to see that
\be
\label{fomal2}
\partial_{\pm}H(y,\xi,t) &=&-\left(\frac{y}{\xi}\mp1\right)\frac{\partial H(y,\xi,t)}{\partial y}\\
&+&\sum^{\infty}_{n=1,3...}\sum^{n+1}_{l=0,2...}
B_{nl}(t)\theta\left(1-\frac{x^{2}}{\xi^{2}}\right)\left(1-\frac{x^{2}}{\xi^{2}}\right)C^{3/2}_{n}\left(\frac{x}{\xi}\right)\frac{\partial
P_{l}\left(\frac{1}{\xi}\right)}{\partial \xi}.
\nonumber
\ee
Substituting this expression into Eq.~(\ref{elementaryTW3WW}) we obtain the partial wave representation of the twist-3 amplitude:
\be
\label{PWAtwist3}
 A^{\rm {tw3}}(\xi,t)=A^{{\rm tw2}}(\xi,t)+2\sum^{\infty}_{n=1,3...}\sum^{n+1}_{l=0,2...}B_{nl}(t)
 \xi\frac{\partial
P_{l}(\frac{1}{\xi})}{\partial \xi}\frac{1}{(n+1)(n+2)}.
\ee
Here $A^{{\rm tw2}}$ is the twist-2 DVCS amplitude given by Eqs.~(\ref{IM},\ref{RE}), so that the new ingredient is the second term in above expression.
For that term we introduce the following notation:

\be
\label{aplus}
A_+=2\sum^{\infty}_{n=1,3...}\sum^{n+1}_{l=0,2...}B_{nl}(t)
 \xi\frac{\partial
P_{l}(\frac{1}{\xi})}{\partial \xi}\frac{1}{(n+1)(n+2)}\, ,
\ee
and perform resummation of partial waves for it. To perform this resummation we, following Ref.~\cite{MaxAndrei}, represent the coefficients
$B_{nl}(t)$ as the Mellin moments of forward-like functions $Q_{2\nu}(x,t)$:

\be
B_{n,n+1-2\nu}(t)=\int_0^1 dx\ x^n\ Q_{2\nu}(x,t)\,.
\ee
Substituting this equation to (\ref{aplus})
changing sequence of summation and integration we obtain:
\begin{eqnarray}
A_{+}(\xi,t)
&=&  2 \sum^{\infty}_{\nu=0}\xi\frac{\partial}{\partial \xi}
\int_{0}^{1}\frac{dx}{x} x^{k}S_{2\nu}(x,t)\sum^{\infty}_{l=2,4..}
x^{l} P_{l}\bigg(\frac{1}{\xi}\bigg),
\label{suml}
\end{eqnarray}
where introduced functions  satisfy following equation
$x^{2}\frac{d^{2} S_{2\nu}(x,t)}{dx^{2}}=Q_{2\nu}(x,t)$ with conditions
$S_{2\nu}(1,t)=\frac{d S_{2\nu}(x,t)}{dx}=0$. It is easy to see that a
solution of the equation has the form of Mellin convolution:
$$S_{2\nu}(x,t)=\int^{1}_{x}\frac{dz}{z}\left(1-\frac x z \right)Q_{2\nu}(z,t).$$
 Eventually, performing the summation over $l$ in Eq.~(\ref{suml}) we obtain the following expression for amplitude $A_{+}(\xi,t)$:
 \begin{eqnarray}
 \label{aplusIMRE}
 {\rm Im} A_{+}(\xi,t)&=&
 \xi\frac{\partial}{\partial \xi}\int^{1}_{\frac{1-\sqrt{1-\xi^{2}}}{\xi}}\frac{dx}{x} {N_S}(x,t)\ \frac{1
 }{\sqrt{\frac{2x}{\xi}-1-x^{2}}},\\
 \nonumber
 {\rm Re}A_{+}(\xi,t)&=&
\xi\frac{\partial}{\partial \xi}\int_0^{\frac{1-\sqrt{1-\xi^2}}{\xi}} \frac{dx}{x} {N_S}(x,t)\ \Biggl[
\frac{1}{\sqrt{1-\frac{2 x}{\xi}+x^2}} + \frac{1}{\sqrt{1+\frac{2
x}{\xi}+x^2}}-\frac{2}{\sqrt{1+x^2}}
\Biggr]  \\
&+&\xi\frac{\partial}{\partial \xi}\int^1_{\frac{1-\sqrt{1-\xi^2}}{\xi}} \frac{dx}{x} {N_S}(x,t)\
\Biggl[ \frac{1}{\sqrt{1+\frac{2
x}{\xi}+x^2}}-\frac{2}{\sqrt{1+x^2}} \Biggr].\nonumber
 \end{eqnarray}
Here in analogy to GPD-quintessence function $N(x,t)$ (\ref{QF}) we introduced its counterpart for twist-3 GPDs:

\be
\label{QF3}
{N_S}(x,t)=\sum_{\nu=0}^\infty x^{2\nu}\ S_{2\nu}(x,t)=\int^{1}_{x}\frac{dz}{z}\left(1-\frac x z \right)\ \sum_{\nu=0}^\infty x^{2\nu}\ Q_{2\nu}(z,t)\, .
\ee
We see that at twist-3 level the function ${N_S}(x,t)$ provides us with new additional information about forward-like functions $Q_{2\nu}(x,t)$.
Therefore it is important to perform Abel tomography for the twist-3 DVCS amplitude.

\section*{\normalsize \bf Abel tomography and dispersion relations for twist-3 DVCS amplitude }
\noindent
Now we can obtain the function ${N_S}(x,t)$ if we know the
twist-3 DVCS amplitude. We remind that the twist-3 DVCS amplitude given by Eq.~(\ref{elementaryTW3}) in WW approximation
can be represented as:

\be
A^{{\rm tw3}}(\xi,t)=A^{{\rm tw2}}(\xi,t)+A_+(\xi,t) \, ,
\ee
where twist-2 amplitude $A^{{\rm tw2}}(\xi,t)$ is given by Eq.~(\ref{IM},\ref{RE}) in terms of GPD-quintessence function $N(x,t)$
and the amplitude $A_+(\xi,t)$ by the Eq.~(\ref{aplusIMRE}) in terms of function ${N_S}(x,t)$ (\ref{QF3}).
Eq.~(\ref{aplusIMRE}) can be brought to Abel type integral equation with help of Joukovski transformation and one can apply the
Abel tomography described in the first section to the twist-3 case. The result is the following:

\be
\label{nani}
{N_S}(x,t)=\frac{1}{\pi}\frac{(1-x^{2})}{\sqrt{1+x^{2}}}\int^{1}_{\frac{2x}{1+x^{2}}}\frac{d\xi}{\sqrt{\xi}}\frac{1}{\sqrt{\xi-\frac{2x}{1+x^{2}}}}
\left\{{\rm Im}A^{{\rm tw2}}(\xi,t)-{\rm Im}A^{{\rm tw3}}(\xi,t)\right\}.
\ee
Measuring twist-2 and twist-3 DVCS amplitudes (they can be separated studying the azimuthal angle dependence of various DVCS asymmetries) we can
access the function ${N_S}(x,t)$ which provides additional information about forward-like functions and GPDs.
Using the expression for ${\rm Im} A^{\rm tw2}$  (\ref{IM}) in terms of the quintessence function $N(x,t)$ we can rewrite the Eq.~(\ref{nani})
in the following form:

\be
N_S(x,t)=\frac{1-x^2}{\sqrt{1+x^2}}\ \left(\int_x^1 \frac{dy}{y}\ \frac{N(y,t)}{\sqrt{1+y^2}}-\frac{1}{\pi}\
\int^{1}_{\frac{2x}{1+x^{2}}}\frac{d\xi}{\sqrt{\xi}}\ \frac{{\rm Im}A^{{\rm tw3}}(\xi,t)}{\sqrt{\xi-\frac{2x}{1+x^{2}}}} \right)
\ee

It was shown in Ref.~\cite{tomography} that the dual parameterization representation for the twist-2 DVCS amplitude (\ref{IM},\ref{RE}) automatically satisfy the
once subtracted dispersion relation:

\be
{\rm Re} A^{{\rm tw2}}(\xi,t)= 2 D(t)+\frac{1}{\pi}\ vp \int_0^1 d\zeta \ {\rm Im} A^{{\rm tw2}}(\zeta,t) \left(\frac{1}{\xi-\zeta}-
\frac{1}{\xi+\zeta} \right)\, ,
\label{DR}
\ee
with the subtraction constant given by the D-form factor (\ref{DFF}).  This result was obtained recently in Refs.~\cite{Teryaev,DiehlIvanov} by independent methods.
We see that the dual parameterization automatically ensures the dispersion relations for the amplitudes. Now we can repeat the derivation of the dispersion relations for
the twist-3 amplitude. To finish this, one has to substitute the expression for ${N_S}(x,t)$ (\ref{nani}) into Eq.~(\ref{aplusIMRE}) for the twist-3 real part of the DVCS amplitude.
After simple calculations we obtain once subtracted dispersion relation for the twist-3 DVCS amplitude:

\be
{\rm Re} A^{{\rm tw3}}(\xi,t)= 2 D(t)+\frac{1}{\pi}\ vp \int_0^1 d\zeta \ {\rm Im} A^{{\rm tw3}}(\zeta,t) \left(\frac{1}{\xi-\zeta}-
\frac{1}{\xi+\zeta} \right)\, .
\label{DR3}
\ee
We see that for the twist-3 amplitude the subtraction constant is given by the D-form factor,
the same as for the twist-2 amplitude.

\section*{\normalsize \bf Conclusions}
\noindent We extended the dual parameterization to the case of the
twist-3 contributions to the DVCS amplitude. It is demonstrated
that the twist-3 DVCS amplitude in the Wandzura-Wilczek
approximation is determined by single function $N_S(x,t)$--the
twist-3 analogue of the GPD quintessence function $N(x,t)$--which
gives complementary information about forward -like functions and
GPDs. We proved dispersion relations for the twist-3 amplitude. It
turned out that the subtraction constant for this case is the same
as for the twist-2 case, and given by the D - form factor.

We presented equation which directly and unambiguously express GPD quintessence functions $N(x,t)$ and $N_S(x,t)$
in terms of twist-2 and twist-3 DVCS amplitudes respectively. Relation of these equations to the Abel transform
tomography is elaborated. Amplitudes of hard exclusive processes can be viewed as ``photographs''
of some object with axial symmetry in plane of variables $X=\sqrt{\frac{2 x}{1+x^2}-\xi}$ and $Y=\sqrt\xi$. It remains to decipher
the deep meaning of these variables and corresponding axial symmetry.

\section*{\normalsize\bf Acknowledgements}
We are thankful to I.~Anikin, N.~Kivel, D.~M\"uller, K.~Semenov-Tian-Shansky, M.~Vanderhaeghen and A.~Vladimirov for many
valuable discussions.
The
work is supported
by the Sofja Kovalevskaja Programme of the Alexander von Humboldt
Foundation, and by  the Deutsche Forschungsgemeinschaft,
 the Heisenberg--Landau Programme grant 2007,
 and the Russian Foundation for Fundamental Research
 grants No. 06-02-16215 and 07-02-91557.

\end{document}